\documentclass{epl}
\title{Low-Temperature Excitations of Dilute Lattice Spin Glasses} 
\author{Stefan Boettcher\footnote{http://www.physics.emory.edu/faculty/boettcher}}
\institute{Physics Department, Emory University, Atlanta, Georgia
30322, USA}


\begin{document}
\maketitle 

\begin{abstract} 
A new approach to exploring low-temperature excitations in
finite-dimensional lattice spin glasses is proposed. By focusing on
bond-diluted lattices just above the percolation threshold, large
system sizes $L$ can be obtained which lead to enhanced scaling
regimes and more accurate exponents. Furthermore, this method in
principle remains practical for any dimension, yielding exponents that
so far have been elusive.  This approach is demonstrated by
determining the stiffness exponent for dimensions $d=3$, $d=6$ (the
upper critical dimension), and $d=7$.  Key is the application of an
exact reduction algorithm, which eliminates a large fraction of spins,
so that the reduced lattices never exceed $\sim10^3$ variables for
sizes as large as $L=30$ in $d=3$, $L=9$ in $d=6$, or $L=8$ in
$d=7$. Finite size scaling analysis gives $y_3=0.24(1)$ for $d=3$,
significantly improving on previous work.  The results for $d=6$ and
$d=7$, $y_6=1.1(1)$ and $y_7=1.24(5)$, are entirely new and are
compared with mean-field predictions made for $d\geq6$.
\end{abstract} 

\section{Introduction}
In this Letter we propose to study low-energy excitations in
Edwards-Anderson spin glasses using {\it bond-diluted} lattices. All
previous studies of such excitations have been hampered by
less-than-adequate scaling behavior that can be obtained with the
limited systems sizes $L$ accessible on undiluted
lattices. Consequently, important questions regarding the spin-glass
state in finite dimensional systems remain unresolved~\cite{KM,PY2}.

A diluted lattice exhibits less frustration on short distances, and it
would appear that the trade-off between potentially larger, yet less
frustrated lattices should not improve the attainable scaling
regime. Here, we find for the case of the ground-state
stiffness~\cite{SY,Kirkpatrick,F+H} of $\pm J$ spin glasses on diluted
hyper-cubic lattices that instead scaling corrections {\it diminish}
with the dilution. Hence, in combination with finite-size
scaling, a dramatically increased scaling window is obtained. This
improves the value for the stiffness exponent from the previously
accepted value of at best $y_3=0.2(1)$~\cite{BM,Hy3d,PY} to our new
value of $y_3=0.24(1)$, which may make the difference between merely
knowing that a positive $y_3$ exists and potentially using exponents
to distinguish theories~\cite{PY2}.

As a further benefit, the relevant bond-density regime is located just
above the percolation transition, at which the mean connectivity for
diluted lattices barely exceeds unity, invariably, for {\it any}
dimension $d$. Accordingly, using exact graph reduction methods, we
can study excitations in all dimensions with system sizes $L$ that are
beyond the scope of undiluted lattices. For instance, reducing
lattices with up to $9^6$ spins, we obtain an entirely new result for
the upper critical dimension $d=6$, $y_6=1.1(1)$. Similarly, using
lattices with up to $7^8$ spins, we obtain $y_7=1.24(5)$, which
suggests that the stiffness exponent increases above the upper
critical dimension.  For results in $d=4$ and 5, see
Ref.~\cite{defect_long}.

On one hand, an increase of $y_d$ with $d$ would appear to be
consistent with replica theory (RSB) predictions~\cite{AMY}.  At least
for $d\geq6$ bulk spins should dominate energy fluctuations induced at
the boundary (see Eq.~(20) in Ref.~\cite{AMY}), implying $y_d=d\mu$,
where $\mu$ is the exponent governing energy fluctuations in the
(infinite-dimensional) Sherrington-Kirkpatrick model. On the other
hand, Ref.~\cite{AMY} also conjectures $\mu=1/4$, i.~e. $y_6^{\rm
RSB}=1.5$ and $y_7^{\rm RSB}=1.75$, well above the results presented
here.

\section{Measuring Stiffness}
Stiffness refers to the ability of a spin system in its ordered state
to resist the nucleation of domains of overturned spins. The creation
of such domains may entail an energetic penalty for forming an
interface. The more ordered the state, the higher the energetic
barrier and the more ``stiff'' the resistance. In turn, if the
creation of such an interface is accomplished with no or reduced
penalty for increasing linear size $L$ of the domain, one may conclude
that the system is in a high-temperature state and poses no
resistance against fluctuations spreading through the
system. Thus, the stiffness exponent $y$ (often labeled $\theta$) is a
fundamental quantity to characterize the low-temperature state of a
spin glass~\cite{F+H,FH,BM1}, and its value is frequently
relied upon~\cite{PY2,BKM}. For instance, Ref.~\cite{BKM} arrives at
conclusions excluded by the more precise value presented here.

The stiffness exponent is typically measured via the ``defect'' energy
$\Delta E$ of an interface induced in a system of size $L$ by swapping
between periodic and anti-periodic boundary conditions along one
spatial direction. In a spin glass the interface of such a growing
domain can take advantage of already-frustrated bonds to grow at a
reduced or even vanishing cost.  The width $\sigma$ of the
distribution $P(\Delta E)$ provides a measure for the energetic cost
of growing a domain of overturned spins. To wit,
\begin{eqnarray}
\sigma(\Delta E)\sim L^y,
\label{yeq}
\end{eqnarray}
where $L$ here refers to the size of a system with an inverted
boundary condition. Clearly, it must be $y\leq d-1$, and a bound of
$y\leq (d-1)/2$ has been proposed for spin
glasses~\cite{FH}. Particularly, ground states of systems with
$y\leq0$ are unstable with respect to spontaneous fluctuations, which
can grow at no cost, like in the case of the one-dimensional
ferromagnet where $y=d-1=0$. Such a system does not manage to attain
an ordered state for any finite temperature. Conversely, a positive
$y$ at $T=0$ indicates a finite-temperature transition into
an ordered regime while its value is a measure of the stability of the
ordered state.

Due to its importance, there have been many attempts to approximate
stiffness exponents in finite-dimensional spin
glasses~\cite{SY,Kirkpatrick,BC,BM,PY,Hy3d,Hd4,CBM,HBCMY,MKpaper},
using transfer matrix, optimization, or renormalization group
techniques. It has been argued long ago that $y<0$ for $d\leq2$ and
$y>0$ for $d\geq3$~\cite{SY,BM}. Only recently, the stiffness exponent
for $d=2$, below the lower critical dimension, has been determined to
considerable accuracy, $y_2=-0.282(2)$~\cite{HBCMY,CBM}. There has
been little progress for $d\geq3$, despite significant increases in
computational power. In $d=3$, the accepted value so far has been
$y_3\approx0.19$~\cite{BM,Hy3d}, although there have been
investigations recently pointing to a larger value, such as
$0.23$~\cite{PY} or $0.27$~\cite{CBM}. All of these studies are based
on fitting power-laws over exceedingly narrow scaling windows at
relatively small system sizes, $4\leq L\leq10$, causing large uncertainties.
Our value in $d=3$ is at the upper end of previous estimates and
amazingly close to (but distinct from) the Migdal-Kadanoff prediction,
$y_3^{\rm MK}=0.25546(3)$~\cite{MKpaper}.

To understand the shortcomings of previous investigations, it is
important to appreciate the complexity of the task: Most numerical
studies are based on sampling the variance $\sigma(\Delta E)$ of the
distribution of defect energies $P(\Delta E)$ obtained via inverted
boundary conditions (or variants thereof~\cite{CBM}). Thus, for an
Ising spin glass with periodic boundaries, an instance of fixed,
random bonds is generated, its ground-state energy determined, then
all bonds within a hyperplane have their sign reversed and the
ground-state energy is determined again. The defect energy $\Delta E$
is the often-minute difference between those two 
energies. Many instances of a given size $L$ have to be generated to
sample the distribution of $\Delta E$ and its width $\sigma(\Delta E)$
accurately. Finally, $\sigma(\Delta E)$ has to be fitted to
Eq.~(\ref{yeq}) for a large and asymptotic range of $L$.

Even small errors in the energy for either boundary condition, by way
of their subtraction, may soon lead to extreme inaccuracies in
$P(\Delta E)$. While for $d\leq2$ efficient algorithms exist to
determine ground state energies exactly, and large system sizes can be
obtained~\cite{HBCMY,CBM}, for $d\geq3$ no such algorithm exists: The
minimization problem is NP-hard~\cite{Barahona} with the cost of exact
algorithms rising faster than any power of $L$. Hence, the values
quoted previously for $y_3$ were either based on small systems,
$L\leq4$~\cite{BM}, or on elaborate heuristic methods with $L\leq10$
that lead to statistical and systematic errors~\cite{Hy3d,PY}.

\section{Reduction Algorithm}
To overcome those limitations, we propose to increase system size $L$
{\it without} increasing the optimization problem by considering
reduced, bond-diluted lattices. As long as the bond density $p$ is
sufficiently above the bond-percolation threshold $p_c$, the dominant
cluster is effectively compact so that the asymptotic scaling behavior
expressed in Eq.~(\ref{yeq}) -- and the stiffness exponent $y$ -- is
{\it independent} of the bond density $p$~\cite{BF,MKpaper}. Furthermore, 
we have developed a new, exact
algorithm, that is capable of drastically reducing the size of
sparsely connected spin glass systems, leaving a much reduced graph
whose ground state can be approximated with great accuracy.

We will describe the reduction algorithm in more detail
elsewhere~\cite{eo_rg}, including its ability to compute the entropy
density and overlap for sparse spin glass systems (see
also~\cite{MKpaper}). We focus here exclusively on the reduction rules
for the energy at $T=0$. These rules apply to general Ising spin glass
Hamiltonians
\begin{eqnarray}
H=-\sum_{<i,j>}\,J_{i,j}\,x_i\,x_j
\label{Heq}
\end{eqnarray}
with {\it any} bond distribution $P(J)$, discrete or continuous, on
arbitrary sparse graphs. For convenience, we use a $\pm J$
distribution on bond-diluted hyper-cubic lattices here.

The reductions effect both spins and bonds, eliminating recursively
all zero-, one-, two-, and three-connected spins. These operations
eliminate and add terms to the expression for the Hamiltonian in
Eq.~(\ref{Heq}), but leave it form-invariant. Offsets in the energy
along the way are accounted for by a variable $H_o$, which is {\it
exact} for a ground state configuration.

{\it Rule I:} An isolated spin can be ignored entirely.

{\it Rule II:} A one-connected spin $i$ can be eliminated, since its
state can always be chosen in accordance with its neighboring spin $j$
to satisfy the bond $J_{i,j}$. For its energetically most favorable
state we adjust $H_o:=H_o-|J_{i,j}|$ and eliminate the term
$-J_{i,j}\,x_i\,x_j$ from $H$.

{\it Rule III:} A double bond, $J_{i,j}^{(1)}$ and $J_{i,j}^{(2)}$,
between two vertices $i$ and $j$ can be combined to a single bond by
setting $J_{i,j}= J_{i,j}^{(1)}+J_{i,j}^{(2)}$ or be eliminated
entirely, if the resulting bond vanishes. This operation is very
useful to lower the connectivity of $i$ and $j$ at least by
one.

{\it Rule IV:} Replacing a two-connected spin $i$ between some spins $1$ and
$2$, the graph obtains a new bond $J_{1,2}$, and acquires an offset
$H_o:=H_o-\Delta H$, by rewriting
in Eq.~(\ref{Heq})
\begin{eqnarray}
&x_i(J_{i,1}x_1+J_{i,2}x_2)\leq\left|J_{i,1}x_1+J_{i,2}x_2\right|
=J_{1,2}x_1x_2+\Delta H,&\nonumber\\
\label{2coneq}
&J_{1,2}=\frac{1}{2}\left(\left|J_{i,1}+J_{i,2}\right|-\left|J_{i,1}-J_{i,2}\right|\right),\quad
\Delta
H=\frac{1}{2}\left(\left|J_{i,1}+J_{i,2}\right|+\left|J_{i,1}-J_{i,2}\right|\right).&
\end{eqnarray}

{\it Rule V:} A three-connected spin $i$ can be reduced via a
``star-triangle'' relation, see Fig.~\ref{startri}:
\begin{eqnarray}
\label{3coneq}
&J_{i,1}\,x_i\,x_1+J_{i,2}\,x_i\,x_2+J_{i,3}\,x_i\,x_3
\leq
J_{1,2}\,x_1\,x_2+J_{1,3}\,x_1\,x_3+J_{2,3}\,x_2\,x_3+\Delta H,&\nonumber\\
&J_{1,2}=-A-B+C+D,\quad J_{1,3}=A-B+C-D,\quad J_{2,3}=-A+B+C-D,&\nonumber\\
&\Delta H=A+B+C+D, \qquad
A=\frac{1}{4}\left|J_{i,1}-J_{i,2}+J_{i,3}\right|,&\\
& B=\frac{1}{4}\left|J_{i,1}-J_{i,2}-J_{i,3}\right|,\quad
C=\frac{1}{4}\left|J_{i,1}+J_{i,2}+J_{i,3}\right|,\quad
D=\frac{1}{4}\left|J_{i,1}+J_{i,2}-J_{i,3}\right|.&\nonumber
\end{eqnarray}
The bounds in Eqs.~(\ref{2coneq}-\ref{3coneq}) become {\it exact} when
the remaining graph takes on its ground state.  Reducing 
higher-connected spins leads to (hyper-)bonds between multiple 
spins, unlike Eq.~(\ref{Heq}).

After a recursive application of these rules, the original lattice
graph is either completely reduced (which is almost always the case
for $p<p_c$), in which case $H_o$ provides the exact ground state
energy already, or we are left with a highly reduced, compact graph in
which no spin has less than four connections. We obtain the ground
state of the reduced graph with the extremal optimization
heuristic~\cite{eo_prl}, which together with $H_o$ provides a very
accurate approximation to the ground state energy of the original
diluted lattice instance.

\begin{figure}
\twofigures[width=5.5cm,angle=-90]{3_con.ps}{semi36pstar.ps}
\caption{``Star-triangle'' relation to reduce a
three-connected spin $x_0$. The new bonds below are obtained in Eq.~(\protect\ref{3coneq}).  }
\label{startri}                                                              
\caption{Log-log plot of the variance $\sigma(\Delta E)$ of the defect
energy as a function of systems size $L$ for various bond fractions
$p>p_c$ in $d=3$ (left) and 6 (right). At small $p$, $\sigma(\Delta
E)$ drops to zero rapidly for increasing $L$, but turns around and
rises for larger $p$, indicative of a nontrivial ordered state at low
$T$. The longest transients in $\sigma(\Delta E)$ mark $p^*$, suggesting $p^*_{d=3}=0.272(1)$ and
$p^*_{d=6}=0.0950(3)$.}
\label{pstarfig}
\end{figure}

\section{Numerical Results}
In Ref.~\cite{BF} it was shown that spin glasses with a discrete bond
distribution on diluted lattices may possess a distinct critical point
$p^*$ in their bond density, which is related to the (purely
topological) percolation threshold $p_c$ of the lattice {\it and} the
distribution of the bond weights $P(J)$. Clearly, no long-range
correlated state can arise below $p_c$. A critical point distinct from
percolation, $p^*>p_c$, emerges when such a correlated state above
$p_c$ remains suppressed due to collaborative effects between
bonds~\cite{BF} (see {\it Rule III}). Thus, to observe any glassy
properties on a dilute lattice, we have to ascertain $p>p^*$. In
Ref.~\cite{MKpaper}, we were able to locate $p^*$ for the
Migdal-Kadanoff lattice, by using the defect energy scaling from
Eq.~(\ref{yeq}): For all $p>p^*$ the stiffness exponent $y$ eventually
took on its $p=1$ value, while for any $p<p^*$ defect energies
diminished rapidly for increasing $L$.

In each dimension (see Ref.~\cite{defect_long}), we have run the above
algorithm on a large number of graphs (about $10^5-10^6$ for each $L$
and $p$) for $p$ increasing from $p_c$ in small steps. For each given
$p$, $L$ increased until it was clear that $\sigma(\Delta E)$ would
either drop or rise for good. In this way, we bracket-in $p^*$, as
shown in Figs.~\ref{pstarfig} for $d=3$ and 6. Tab.~\ref{datatable}
summarizes those results.

\begin{figure}
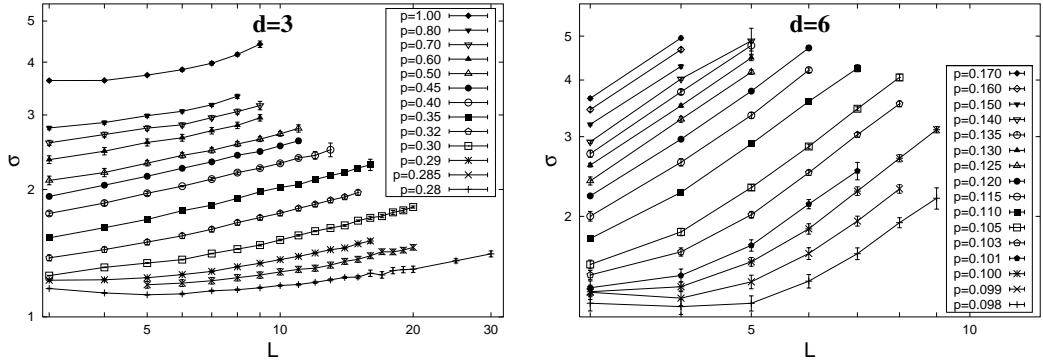

\twoimages[width=7cm]{3d_raw.eps}{6d_raw.eps}
\caption{Log-log plot of the width $\sigma$ of the defect energy
distribution as a function of system size $L$ for $d=3$ (top) and 6
(bottom).  The data is grouped into sets (connected by lines)
parameterized by the bond density $p$. Most sets show a distinct
scaling regime as in Eq.~(\protect\ref{yeq}) for a range of $L$ above
finite scaling corrections but below failing heuristic accuracy.  }
\label{rawdefectplot}
\end{figure}

While the value of $p^*$ is distribution-dependent, we merely had to
establish a bond fraction beyond which we would expect Eq.~(\ref{yeq})
to hold. Then we can conduct numerical experiments to extract the
asymptotic scaling of $\sigma(\Delta E)$ for conveniently chosen
$p$. For each choice of $L$ and $p$, we have sampled the defect energy
distribution with at least $N\geq10^5$ instances, then determined its
variance $\sigma(\Delta E)$. Throughout, the distribution of $\Delta
E$ is approximately normal, giving rise to an error bar
$\propto1/\sqrt{N}$. In Fig.~\ref{rawdefectplot}, we plot all the
data for each dimension simply according to Eq.~(\ref{yeq}), on a
logarithmic scale. For most sets of graphs, a scaling regime (linear
on this scale) is visible. Yet, various deviations from scaling can be
observed. Clearly, each sequence of points should exhibit some form of
finite size corrections to scaling for smaller $L$. For large $L$, the
inability to determine defect energies correctly, will inevitably lead
to a systematic bias in $\sigma$. Some data sets did not exhibit any
discernible scaling regime, most notably our set for the
undiluted lattice in $d=3$. This resembles the
observation of Refs.~\cite{DM}, which found long transients in
similar studies on undiluted $d=2$ or Migdal-Kadanoff lattices.

\begin{figure}
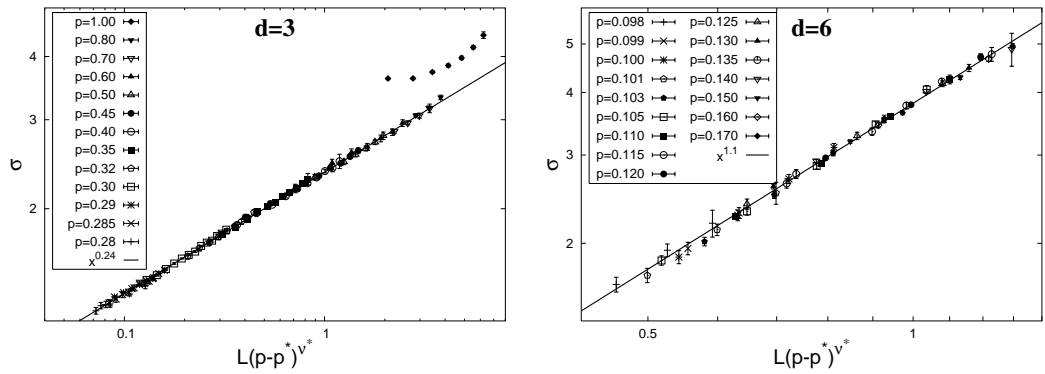

\twoimages[width=7cm]{3dscal.eps}{6dscal.eps}
\caption{Scaling plot of the data from
{}Figs.\protect\ref{rawdefectplot} for $\sigma$, fitted to the
functional form in Eq.~(\protect\ref{fiteq}) as a function of the
scaling variable $x=L(p-p^*)^{\nu^*}$ for $d=3$ (top) and 6
(bottom). Data that fell outside of the scaling regime in each set of
Figs.~\protect\ref{rawdefectplot} was cut. The straight line in each
case represents the fit according to Eq.~(\protect\ref{fiteq}) for the
optimal collapse of the data, which provides an accurate determination
of the stiffness exponent $y$ in each dimension. For $d=3$, we have
also included the data for $p=1$, which does not appear to have a
scaling regime. }
\label{scaldefectplot}
\end{figure}

To obtain an optimal scaling collapse of the data, we focus on the
data inside the scaling regime for each set. To this end, we chose for
each data set a lower cut in $L$ by inspection. An appropriate
high-end cut is introduced by eliminating all data points for which
the remainder graph had a typical size of $>700$ spins; at that point
the EO heuristic (within the supplied runtime) seems to fail in
determining defect energies with sufficient accuracy. All the
remaining data points for $L$ and $p$ are fitted to a four-parameter
scaling form,
\begin{eqnarray}
\sigma(\Delta E)\sim {\cal Y}_0\,
\left[L\left(p-p^*\right)^{\nu^*}\right]^y\quad (L\to\infty),
\label{fiteq}
\end{eqnarray}
suggested by Ref.~\cite{BF}. Unfortunately, we have a-priori no knowledge of
corrections for small systems, making the low-$L$
cut on the data a necessity. The fitted values for the fitting
constants ${\cal Y}_0$, $\nu^*$, and $y$ are listed in
Tab.~\ref{datatable}. Using the parameters of this fit, we re-plot only
the data from the scaling regime in
Figs.~\ref{scaldefectplot}.

\begin{figure}
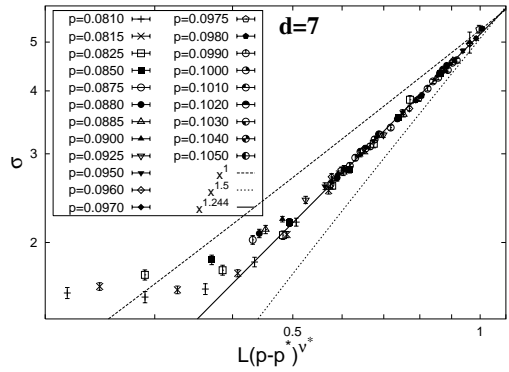

\onefigure[width=7cm]{7dscal.eps} s
\caption{Scaling plot of the data for $\sigma$ in $d=7$ dimensions,
fitted to the functional form in Eq.~(\protect\ref{fiteq}) as a
function of the scaling variable $x=L(p-p^*)^{\nu^*}$. As before, data
outside of the scaling regime, although plotted here, was disregarded
in the fit. The straight line represents the fit according to
Eq.~(\protect\ref{fiteq}) for the optimal collapse of the data. It
seems to exclude a simple (half-)integer value for $y_7$, such as
unity or 3/2 (dashed lines). }
\label{scaldefectplotd7}
\end{figure}

We have also considered the case of $d=7$. The implementation of our
algorithm is not yet capable of handling lattice graphs with much more
than $10^6$ spins, limiting attainable sizes to $L\leq8$, which was
insufficient to determine $p^*$ directly (as in Fig.~\ref{pstarfig})
as well as the low end of the scaling variable
$x=L(p-p^*)^{\nu^*}$. Yet, we have computed about as many data points
as for $d=6$ (see ``DoF'' in Tab.~\ref{datatable}), as shown in the
scaling plot in Fig.~\ref{scaldefectplotd7}. Solid scaling extends
over nearly half a decade, yielding $y_7=1.24(5)$. Assuming that the
data presented in Fig.~\ref{scaldefectplotd7} is asymptotic, any value
for $y_7$ outside of its error bars has a ``goodness of fit''
$Q$~\cite{Press} below 0.001, and $Q<10^{-100}$ for 1 or
$3/2$. Generally, $y_d$ appears to be rising with $d$ above the upper
critical dimension, but with values and at a rate below replica theory predictions~\cite{AMY}.

\begin{table}
\caption{List of the values for the critical bond-density $p^*$ (from
Fig.~\protect\ref{pstarfig} for $d=3,6$), and the fitted values according to
Eq.~(\protect\ref{fiteq}) of the correlation-length exponents $\nu^*$,
the surface tensions ${\cal Y}_0$, and the stiffness exponents $y$,
followed by the ``goodness of fit,'' $Q$~\protect\cite{Press}, and the
degrees of freedom (DoF) used in the fit. Also listed are values for $d=4$
and 5 from Ref.~\protect\cite{defect_long} and the fitted values for $d=7$.}
\begin{tabular}{r|lllll}
\hline\hline 
$d$ & $p^*$ & $\nu^*$ & ${\cal Y}_0$ & $y$ & $Q$(DoF)\\ 
\hline 
3 & 0.272(1)  & 1.17 & 2.37 & 0.239 & 1.00(92) \\ 
4 & 0.1655(5) & 0.60 & 2.43 & 0.610 & 0.00(47) \\ 
5 & 0.1206(2) & 0.50 & 3.05 & 0.876 & 0.86(48) \\ 
6 & 0.0950(3) & 0.44 & 3.87 & 1.103 & 0.02(46)  \\ 
7 & 0.080(1)  & 0.40 & 5.18 & 1.244 & 1.00(55)  \\ 
\hline\hline
\end{tabular}
\label{datatable}
\end{table}

\acknowledgments
I would like to thank A. Percus, F. Krzakala, T. Aspelmeier, and R. Palmer for helpful discussions,
and our IT staff for providing access to our student computing
lab. This work was supported by NSF grant DMR-0312510.

\end{document}